\documentclass[manuscript]{aastex}

\slugcomment{\bf Revised Paper of February 25, 2015}

\shorttitle{Enigmas from the SDSS DR7 White Dwarf Catalog}
\shortauthors{Liebert, Ferrario, Wickramasinghe, Smith}

\begin{document}

\title{{Enigmas from the Sloan Digital Sky Survey DR7 Kleinman White Dwarf 
Catalog}}

\author{James Liebert\altaffilmark{1}, Lilia Ferrario\altaffilmark{2},
Dayal T. Wickramasinghe\altaffilmark{2}, Paul S. Smith\altaffilmark{1} }

\altaffiltext{1}{Steward Observatory, University of Arizona, Tucson AZ
85721; jamesliebert@gmail.com, psmith@as.arizona.edu}

\altaffiltext{2}{Mathematical Sciences Institute, Australian National
University, ACT 0200, Australia; Lilia.Ferrario@anu.edu.au,
Dayal.Wickramasinghe@anu.edu.au}

\begin{abstract}\

  We report results from a continuation of our searches for high field
  magnetic white dwarfs paired in a detached binary with non
  degenerate companions.  We made use of the Sloan Digital Sky Survey
  DR7 catalog of Kleinman et al. (2013) with 19,712
  spectroscopically-identified white dwarfs.  These include 1,735
  white dwarf plus M dwarf detached pairs (almost 10\% of the Kleinman
  at al.'s list).  No new pairs were found, although we did recover the
  polar (AM~Herculis system) ST\,LMi in a low state of accretion.  With
  the larger sample the original situation reported ten years ago
  remains intact now at a much higher level of statistical
  significance: in the selected SDSS sample, high field magnetic white
  dwarfs are not found in an apparently-detached pairing with an M
  dwarf, unless they are a magnetic CV in a low state of accretion.
  This finding strengthens the case that the fields in the isolated
  high field magnetic white dwarfs are generated by stellar mergers
  but also raises questions on the nature of the progenitors of the
  magnetic CVs.

\end{abstract}


\section{Introduction}

After Data Release 1 (2004) of the Sloan Digital Sky Survey (York et
al. 2000), some 169 magnetic white dwarfs were known with fields
greater than about 2 mega-gauss (MG) from SDSS and other sources.  We
refer to these high field magnetic white dwarfs as MagWDs. In the same
release, some 501 detached binary systems consisting of a white dwarf
(WD) and non-degenerate companion were listed.  The companion was
nearly always an M dwarf.  Some ten years ago Liebert et al. (2005)
noticed a curious fact: The two subsets of MagWDs and detached WD +
dM's did not intersect.  To confirm this, we inspected visually all
MagWDs and WD + dM's.  Indeed, no MagWD had a non-degenerate
companion, and no WD + dM detached pair included a MagWD.  These
results were surprising, considering the fact that some $8\%$ of 
isolated white dwarfs were MagWDs (Wickramasinghe \& Ferrario 2000) 
and led Tout et al. (2008) to propose that the isolated MagWDs
were the result of a stellar merger where the fields were generated
during the merger process, a proposal that has since gained
considerable momentum (Nordhaus et al. 2011, Garcia-Berro et
al. 2012, Wickramasinghe et al. 2014, Briggs et al. 2015).

The SDSS WD + detached M samples include both wide pairs in which the
progenitor of the WD and the dM were evolved without interaction and
pairs that underwent common envelope evolution -- the so-called post
common envelope binaries (PCEBs).  The latter group will evolve into
the Cataclysmic Variables (CVs), semi-detached close binary systems
where a nondegenerate late type companion transfers matter to a white
dwarf.  Now about one quarter of known CVs have white dwarfs with
fields in the range $\sim 1-1,000$\,MG, the so-called Magnetic
Cataclysmic Variables (MagCVs, see the review by Wickramasinghe \&
Ferrario 2000).  Our failure to detect magnetic fields in the 
detached WD + dM sample therefore also raises the question ``where 
are the PCEBs that are the progenitors of the MagCVs?''

\section{The Current Study}

In this paper, we report on the continuation of the same searches
utilizing the much larger Data Release 7 (DR7) of SDSS.  Kleinman et
al. (2013) have compiled a white dwarf catalog, with some 19,712
spectroscopically identified white dwarfs with SDSS spectra.  They
also visually identified 804 MagWDs.  Kepler et al. (2013) also
visually inspected the SDSS spectra and estimated the field strength
and masses of 521 DA MagWDs.  Their work has significantly increased 
the number of known MagWDs.  However, none of them was reported to 
have a companion.  

We visually inspected again all 1,735 WD + dM pairs (1,951 individual
spectra including duplicates), almost 10\% of the WD sample listed by
Kleinman et al. (2013).  Most of these binaries are also catalogued
in Rebassa-Mansergas et al. (2010, 2012).  Our search was motivated by
our knowledge that it would have been easy to miss recognizing the
presence of a magnetic field in a white dwarf in previous such
studies, particularly if the field is either very high, or the field
structure is peculiar, or the spectrum is due to elements other than
hydrogen.  The discovery of such a system would have required
follow-up polarimetric or spectropolarimetric observations for
confirmation, but none was found.

We thought that we had found one case where a moderate field MagWD
might be paired with an M dwarf in a detached binary. The SDSS object,
designated SDSSJ~110539.77+250628.5, appeared to be a detached binary
system with a probable magnetic white dwarf having an dM5-6 secondary
companion.  Despite the high signal-to-noise ratio (SNR) of the
spectrum, the probable magnetic features did not appear as simple
Zeeman splitting of hydrogen lines.  The $H\beta$ transition was
clearly present in absorption, but a clear Zeeman triplet was not
obvious.  A narrow $H\alpha$ emission line, presumably coming from the
late-M secondary, was also present in the spectrum.  To confirm its
magnetic nature polarimetric observations were obtained on several
separate runs in 2014 June (by PSS) on the Catalina Observatory 1.55-m
with the SPOL spectropolarimeter.  These showed that the object was an
AM Herculis system (polar) in a high state of accretion.  It turned
out that the low state SDSS spectrum that had led us to believe that
this was a detached system, had already been published by Szkody et
al. (2009) in a paper on multiple cataclysmic variables observed in
the SDSS.  Here the system had been identified as the well known Polar
ST\,LMi with an orbital period of $P_{\rm orb}$ = 113.9\,minutes and a
field $B$=12\,MG (Ferrario et al. 1993) in a very low state of
accretion. The spectrum that we obtained was similar to that obtained
by Stockman et al (1983). During the orbital period, there is a
$\sim$40 minute ``bright phase'' when the actively-accreting pole is
in view, and the circular polarization is about 15\%.  There is a
$\sim$74 minute ``faint phase'' where the circular polarization is
essentially zero (when the active pole is out of view). A cyclotron
harmonic accounts for the rise in the spectrum at longer wavelengths,
when the bright pole is in view.  We have added nothing new to the
understanding of ST~LMi, other than that its behaviour has not changed
in the last $\sim$30 years.

The system we had found was clearly {\it not} what we were originally
looking for, that is, a detached MagWD + M dwarf system.

\section{Discussion and Conclusions} 

Our visual search for apparently detached WD + dM's from the SDSS DR7 
leads us to the conclusion that there is still {\it no known MagWD 
paired with a nondegenerate companion}.  The sample size for these 
searches is now about {\it four times} as large as it was ten years 
ago, approaching a total of 20,000 objects.  A simple $\chi^2$ test 
assuming one degree of freedom shows that the hypothesis that 
magnetic field and binarity (when paired with K or M dwarfs) are 
independent is now vindicated at the $9\,\sigma$ level. 
Notwithstanding the fact that white dwarfs with super-strong fields 
particularly in spectra of low SNR are difficult to identify so that 
some potential candidates may have been missed, the case is now 
overwhelming that there is essentially {\it no intersection between 
MagWDs and WD + dM detached pairs.}  

The continued absence of detached MagWD and nondegenerate dwarf
pairings reinforces the ideas of Reg\"os \& Tout (1995), Tout et al.
(2008), and Wickramasinghe et al. (2014) that strong fields in
isolated MagWDs are generated by a dynamo mechanism that feeds on the
differential rotation that occurs either during common envelope
evolution of a binary system that results in a merger or during a
double degenerate merger (see also Nordhaus et al. 2011, Garcia-Berro
et al 2012).  Theory suggests that a variety of magnetic field
strengths can result from such mergers, with the field strength
perhaps increasing with the combined mass (see e.g. Tout et al.
2008).  Evidence that the mass increases with MagWD field strength is
discussed in Kepler et al. (2013).  That nearby, highly magnetic WDs
with accurate trigonometric parallaxes are more massive than their
non-magnetic counterparts was argued by Liebert (1988).

We now turn to the Cataclysmic Variables (CVs).  Szkody et al. (2011) 
have identified 285 CVs from the SDSS DR7 release and about $20\%$ 
of these are magnetic.  Our failure to detect detached systems with 
a MagWD again raises the question ``where are the PCEBs that are the 
progenitors of the MagCVs?''

There have been specific searches for PCEBs using SDSS surveys 
augmented by radial velocity measurements (e.g. Rebassa-Mansergas 
et al. 2011, Zorotovic et al. 2011, Parsons et al. 2013a).  These 
studies have shown that roughly one third of the detached WD + dM 
pairs in SDSS are PCEBs.  Zorotovic et al. (2011) listed some $60$ 
PCEBs with well determined orbital periods and WD effective 
temperatures.  The periods range from $0.08-20$ days and effective 
temperatures from $\sim 7,500-60,000$~K.  Magnetic braking and 
gravitational radiation will lose angular momentum from the orbit 
and bring about a third of these systems into contact within a 
Hubble time.  Interestingly, none of these had MagWD companions. 

There is a known class of binary system whose members could be 
identified as PCEB pre-polars.  These were initially referred to as 
LARPs -- Low Accretion Rate Polars (Reimers et al. 1999, Reimers \& 
Hagen 2000, Schwope et al. 2002, Schmidt et al. 2005a, 2007).  They 
are detached binaries in which the MagWD accretes matter at a very 
low rate from a stellar wind that emanates from the nondegenerate 
companion (Webbink \& Wickramasinghe 2005).  The magnetic nature of 
the LARPs was discovered through the detection of cyclotron emission 
humps in their energy distribution from accretion shocks on the 
white dwarf surface that were superposed on the WD + dM stellar 
continuum (e.g. Ferrario, Wickramasinghe \& Schmidt 2005).  Such 
systems are not expected to be picked up in surveys such as the one 
used by us to select normally-looking WD + main sequence stellar 
continua.  In fact the first such systems were discovered in surveys 
aimed at discovering active galactic nuclei! 

Schwope et al. (2009) have summarized the properties of the $9$ known 
LARPs, which were renamed as PREPs (pre-polars) to avoid confusion 
with Polars in low states of accretion.  These systems have short 
orbital periods very similar to but on average marginally larger 
that those of Polars.  Their WDs have temperatures that are typically 
much less than $10,000$~K, and their cooling ages ar estimated to be 
greater than about 200 million years with most having ages of above 
a billion years.  In fact, the WDs in Roche lobe-filling polars tend 
to have somewhat higher tempertures with the difference attributed 
to different amount of accretion heating in the high and low modes 
of accretion.  Schmidt et al. (2005a) argued that there were field 
biases in the PREPs discovered from SDSS that depended on the strength 
and positioning of cyclotron harmonics in the filter bands used in 
the selection process that favored only the discovery of strong field 
($\sim 40-100$MG) systems.  It is reasonable to expect larger numbers of
lower field systems in similar states of accretion, although they 
would be less easy to detect as the cyclotron emission peak moves 
further into the infrared (see Parson et al. 2013b).  

In addition to PREPs discovered through the detection of cyctrotron 
lines, there are also MagWDs such as SDSSJ~121209.31+013627.7 
(Schmidt et al 2005b, Farihi et al. 2008, Burleigh et al. 2006, 
Koen \& Maxted 2006, Debes et al. 2006, Linnel et al. 2010) that 
appear single in SDSS spectra but show the presence of very low 
mass companions when observed further in the infrared (Dunlap et al. 
2013, Breedt et al. 2012).  Although in many of these cases it is 
unclear if they represent PREPs or MagCVs in an extended low state, 
it is likely that some of these systems are also PREPs. 

The known PREPs, due to their low effective temperatures and large 
cooling ages, are clearly in a very late stage of post common 
envelope evolution.  The hotter PREPs (and more generally MagCVs) 
should be found among the PCEBs extracted from SDSS and other 
sources (Zorotovic et al. 2011) which does have many hot WDs. 
However, although some $20\%$ of known CVs are magnetic, their 
birthrate may be significantly smaller that that of nonmagnetic CVs 
because of the expected reduction in magnetic braking in these 
systems (Li et al. 1994, Webbink \& Wickramasinghe 2000) for which 
there is some observational evidence (Townsley \& G\"ansicke 2009, 
Araujo-Betancor et al. 2005).  The absence of hot PREPs may not be 
surprising given the small sample size, and with better statistics 
it may indeed turn out that the PREPs can be identified as the 
progenitors of the MagCVs as was already suggested by Webbink \& 
Wickramasinghe (1995).  

We conclude by noting that the physics of the common envelope phase of
evolution is poorly understood and, not surprisingly, there are no
detailed studies of binary evolution that allows for the generation of
magnetic fields during the common envelope phase.  A prediction of the
Tout et al. (2008) model is that the magnetic fields of the white
dwarfs are generated during the common envelope phase with systems
that merge producing isolated MagWDs and systems that do not merge,
but come closer to contact, being the progenitors of the MagCVs
(Polars and Intermediate Polars).  As our present study has shown, the
stellar merging hypothesis appears to be consistent with all available
observations and has also been supported by recent population
synthesis studies (Briggs et al. 2015).  The question of why the
isolated high field MagWDs on the one hand, and the Polars and
Intermediate Polars on the other, both emerge from binary evolution
with magnetic fields in the same range (1-1,000\,MG) could be related
to the robustness of the dynamo model that invokes differential
rotation for the generation of fields (Wickramasinghe et al. 2014).
These aspects of the model should be testable by observations when a
larger sample of PCEBs that all include pre-MagCVs becomes available.

\acknowledgments

We thank Boris G\"ansicke for drawing our attention to the catalogue
of CVs based on SDSS where the low state spectrum of ST\,LMi had
already been reported.  We also wish to thank Scot Kleinman and the
Referee for helpful suggestions.

\end{document}